\documentstyle[aps,pre,amsfonts,preprint]{revtex}
\begin{document}
\draft
\title{\bf Geometric dynamical observables in rare gas crystals}
\author{Lapo Casetti\cite{INFN}}
\address{Scuola Normale Superiore, Piazza dei Cavalieri 7, 56126 Pisa, Italy}
\author{Alessandro Macchi\cite{Mac}}
\address{Istituto Nazionale di Fisica della Materia (INFM), Unit\`a di Firenze,
Largo E. Fermi 2, 50125 Firenze, Italy}
\date {Novembre 20, 1996}
\maketitle
\begin{abstract}
We present a detailed description of how a differential
geometric approach to Hamiltonian dynamics can be used for determining
the  existence of a crossover between different dynamical regimes  
in a realistic system, a model of a rare gas solid. Such a geometric
approach allows to locate the energy threshold between weakly
and strongly chaotic  regimes, and to estimate the largest Lyapunov
exponent. We show how  standard
methods of classical statistical mechanics, i.e. Monte Carlo simulations,
can be used for our computational purposes. Finally we consider
a Lennard Jones crystal modeling solid Xenon. The value
of the energy threshold turns out to be in excellent agreement with
the numerical estimate based on the crossover between slow and fast
relaxation to equilibrium obtained in a previous work 
by molecular dynamics simulations. 
\end{abstract}

\narrowtext
\section{Introduction \label{intro}}

Generic non-integrable Hamiltonian systems with $N \geq 3$ degrees of freedom 
always have a connected chaotic component in phase space. Moreover, as $N$ is
large, the measure of such a component should be practically 
coincident with the
measure of the whole constant-energy hyper-surface. 
In fact, as discussed in Refs. \cite{PettiniLandolfi,PettiniCerruti,Pettini},
the invariant tori whose existence is predicted by the 
Kol'mogorov-Arnol'd-Moser (KAM) theorem have a positive measure only below
a critical value for the amplitude of the non-integrable part of the Hamiltonian,
and this critical amplitude is estimated to be rapidly decreasing with $N$.
This does not mean that particular cases in which the KAM threshold is 
relevant also in macroscopic systems cannot exist, nevertheless
 it suggests that
such a situation could be hardly generic. 

These facts support the expectation that varying the energy --  or more
precisely the energy density $e=E/N$ which is the physical parameter
as  $N$ is  large and eventually as the limit $N \to \infty$ is taken -- 
one should not observe
any qualitative change in the dynamical behavior
of large non-integrable Hamiltonian systems. 
The dynamics should be completely chaotic at all energies, 
and the only effect of a variation of the energy should be a somewhat trivial 
rescaling of the characteristic instability time scale, 
measured by the inverse of the Lyapunov exponent $\lambda$. 
An example of this kind of behavior is provided
by self-gravitating systems \cite{Nbody}. 

On the contrary, there is now a widely accepted numerical
evidence that -- at least as long as models of nonlinear
coupled oscillators are considered --  there exist qualitatively 
different regimes in the dynamics, which have been referred to as weak
and  strong chaos \cite{PettiniLandolfi,PettiniCerruti}. In the strongly
chaotic regime which corresponds to the above sketched scenario, 
fast phase-space 
mixing is observed  regardless of the initial conditions.
At variance, in correspondence
of weak chaos one can observe very long mixing times with non-equilibrium
initial conditions, and the details of the dynamics are strongly influenced
by the choice of the initial conditions. Moreover, at least on finite time
scales, the dynamics appears as globally recurrent, as is probed by the
probability distribution of single-particle autocorrelation functions
\cite{Isola}. This effect was observed 
in several numerical simulations \cite{Bocchieri,Livi_old} inspired by the
results of the celebrated numerical experiment by Fermi,
Pasta and Ulam (FPU) \cite{FPU} where the expected equipartition of energy 
among normal modes was not observed in a chain of linear oscillators
coupled by a weak anharmonicity.

The transition between weak and strong chaos is rather sharp when detected
looking at non-equilibrium properties, i.e. observing the time behavior
of observables which depend on the choice of a particular dynamical 
initial condition in which the system is far from thermodynamic equilibrium.
Examples of these observables are the relaxation times of several dynamical
observables  to their equilibrium value 
\cite{PettiniLandolfi,PettiniCerruti}, the finite-time values 
of the so-called spectral entropy \cite{Livi_old,Livi_new}, and the already
mentioned probability distributions of single-particle autocorrelation
functions \cite{Isola}. The behavior of these observables neatly
detects a threshold value  $e = e_c$ which marks the transition between
weak and strong chaos, and has been referred to as the 
strong stochasticity threshold (SST)
\cite{PettiniLandolfi,PettiniCerruti,Pettini,Nbody,CasettiPettini}, or as the 
crossover energy (CE)  \cite{3deuro}.
All these observables have the drawback of being
not globally defined, i.e. of depending on a particular choice of the
initial condition, which could in principle depend on $N$. 

It is 
remarkable that also the Lyapunov exponent $\lambda$,   
an observable which is neither
dependent on the initial conditions nor on $N$, and which measures directly
the degree of chaos, marks the SST. In fact
in correspondence of $e_c$ the dependence of $\lambda$ on $e$ has a crossover:
in the strongly chaotic regime one 
finds a power law which can be successfully predicted by
a random matrix approximation (RMA) for the tangent dynamics
\cite{PettiniCerruti}. In this regime the RMA yields good predictions also
for the shape of the whole spectrum
of Lyapunov exponents \cite{EckmannWayne}. At variance, in the weakly chaotic 
case $\lambda$ is still positive -- for nonlinear coupled 
oscillators $\lambda \propto e^2$ --  thus
indicating that chaos is present, but the RMA is no longer able to predict
its dependence on $e$. The breakdown of the RMA is a clue that some global
change has happened in the phase space structure. The drawback of using the
dependence of $\lambda$ on the energy density as a probe of the SST is that
the transition is no longer sharp as in the case of non-equilibrium observables:
at $e_c$, $\lambda(e)$ exhibits only a crossover between different asymptotic
behaviors. 

The deep origin of these puzzling dynamical features of non-integrable
Hamiltonian systems with many degrees of freedom
has at present not yet been understood: 
nevertheless, a recently proposed differential geometrical
approach to Hamiltonian chaos \cite{Pettini} has established a link
between the SST and some major change of the geometric structure 
underlying the dynamics, allowing an operative definition of this threshold
no longer based on the computation of 
time-asymptotic quantities like Lyapunov exponents, but on statistical
averages of geometric observables.
Starting from the results of Ref. \cite{Pettini}
it has been proved the stability of the SST in the thermodynamic
limit $N\rightarrow \infty$, at least for the one-dimensional FPU model
\cite{CasettiPettini}. Moreover this approach provides the basic tools
to obtain a model scalar equation which describes the main features of chaos
being independent of the details of the dynamics \cite{Gaussmodel}.
The present work follows this geometric approach.

It is worth mentioning that very recently another kind of geometric 
approach has been put forward \cite{Casartelli} which is based on
the geometric properties of the trajectories -- seen as curves in 
${\Bbb R}^{2N}$ -- rather than on the properties of the ambient
manifolds. As long as a comparison is possible, the two approaches 
yield perfectly consistent results.

The phenomenology related to the SST has been mainly studied in connection
with one-dimensional models of coupled oscillators (FPU model and lattice
$\varphi^4$ classical model \cite{PettiniLandolfi,PettiniCerruti,Pettini}),
and some numerical results for two-dimensional crystals
\cite{LJ2} strongly suggest that a transition between weak and
strong chaos is present in more realistic systems.
The present work is concerned with the possibility of the existence of
the SST in a realistic system, in the perspective of a possible 
experimental verification of the physical consequences of the transition
between weak and strong chaos. Such an experiment has been recently
proposed \cite{3deuro}, and should verify the existence of a  crossover
between slow and fast relaxation to thermal equilibrium in a rare gas
crystal with diluted impurities.  The far-from-equilibrium 
dynamics of such a system has been studied numerically in Ref. \cite{3deuro},
showing that the crossover occurs at $e \simeq  0.15\varepsilon$, where 
$\varepsilon$ is the depth of the interaction potential well.
The present work is concerned with the equilibrium dynamical properties
of a simpler but closely related system, i.e. a model of a rare  gas solid
without impurities. As discussed in Ref. \cite{3deuro}, the role
played by impurities is crucial to obtain suitable initial conditions
in the non-equilibrium case, but should have only a weak effect on the
global dynamical properties of the system (this fact is confirmed
in a 1-d case \cite{1d}), hence the eventual existence of a SST
in our model system -- apart from being interesting by itself -- could
have a significance in explaining the phenomenology observed in Ref. 
\cite{3deuro}. 

More precisely, the model studied in the present work 
is a system of $n$ point masses 
arranged on a three-dimensional fcc lattice with nearest-neighbor 
Lennard-Jones interaction. The classical dynamics of such a system is a
reasonable approximation of the behavior of a rare gas crystal if 
quantum effects can be neglected, i.e. as long as the temperature is
high enough. Hence the existence of a crossover between weak and strong chaos
in this model could 
have a substantial physical relevance if it occurs at $e$ values 
allowing a classical description of the dynamics, at least as a first
approximation. As we shall see in Sec. \ref{results}, this happens at
least in the case of Xe crystals.

The present work is organized as follows: 
in Sec. \ref{geometry} the geometrical methods which allow a characterization
of the transition between weak and strong chaos are sketched;
these methods were introduced in 
Refs. \cite{Pettini,CasettiPettini} and \cite{Gaussmodel}, 
where all details can be found. The 
model studied is described in Sec. \ref{model}, and the results are
presented and discussed in Sec. \ref{results}. 
Some conclusions are drawn in Sec. 
\ref{conclusions}.

\section{Riemannian geometry and chaotic dynamics}
\label{geometry}

Hamiltonian dynamics can be rephrased in geometrical terms owing to
the fact that the trajectories of a dynamical system with quadratic
kinetic energy can be seen as geodesics of a suitable Riemannian 
manifold. There are several choices for the ambient manifold as well as
for the metric tensor. As already discussed in Ref. \cite{CasettiPettini}
a particularly useful ambient space 
is the enlarged configuration space-time $M\times {\Bbb R}^2$, i.e.
the configuration space $\{q^1,\ldots,q^i,\ldots,q^N\}$ with two additional
real coordinates $q^0$ and $q^{N+1}$. In the following $q^0$ will be identified
with the time $t$. For standard Hamiltonians ${\cal H}=T+V({\bf q})$ where 
$T=\frac{1}{2}a_{ij}\dot q^i\dot q^j$, this manifold, 
equipped with Eisenhart's metric $g_E$, 
has a semi-Riemannian (Lorentzian) structure ($\det g_E = -1$).
The arc-length is given by 
\begin{equation}
ds^2=a_{ij}dq^i  dq^j - 2V({\bf q})(dq^0)^2 + 2dq^0 dq^{N+1}~,
\end{equation}
where both $i$ and $j$ run between $1$ and $N$. Let us restrict to geodesics
whose arc-length parametrization is  affine, 
i.e. $ds^2=2C_1^2 dt^2$; simple algebra shows that the geodesic equations
\begin{equation}
\frac{d^2q^\mu}{ds^2}+\Gamma^\mu_{\nu\lambda}\frac{dq^\nu}{ds}
\frac{dq^\lambda}{ds}=0 ~~~~~\mu,\nu,\lambda=0\ldots N+1~,
\end{equation}
become Newton equations (without loss of 
generality $a_{ij}=\delta_{ij}$ is considered)
\begin{equation}
\frac{d^2q^i}{dt^2}  =  - \frac{\partial V}{\partial q_i} 
\label{eqgeoit} 
\end{equation}
for $i=1\ldots N$, together with two extra equations for  
$q^0$ and $q^{N+1}$ which can be integrated to yield
\begin{mathletters}
\label{excoord}
\begin{eqnarray}
q^0  & = & t \label{q0(t)} \\
q^{N+1} &  = & C_1^2 t+ C_2 - \int_0^t L({\bf q},\dot{\bf q})\,dt 
\label{qN+1(t)}
\end{eqnarray}
\end{mathletters}
where $L({\bf q},\dot{\bf q})$ is the Lagrangian, and $C_1$, $C_2$ are 
real constants.
As stated by Eisenhart theorem \cite{Eisenhart}, the dynamical 
trajectories in configuration space are projections on $M$ of the geodesics of
$(M\times{\Bbb R}^2,g_E)$.

In the geometrical framework, the stability 
of the trajectories is mapped on the stability 
of the geodesics, hence it can be studied by the 
Jacobi equation for geodesic deviation
\begin{equation}
\frac{D^2 J}{ds^2} + R(\dot\gamma, J)\dot\gamma = 0~,
\label{eqJ}
\end{equation}
where $R$ is the Riemann curvature tensor, $\dot\gamma$ is the velocity
vector along the reference geodesic $\gamma(s)$,  $D/ds$ is the covariant
derivative and $J$,
 which measures the deviation between nearby
geodesics, is referred to as the Jacobi field. 
The stability -- or instability -- of the dynamics, and thus
deterministic chaos, originates from the curvature properties 
of the ambient manifold. In local coordinates, Eq. \ref{eqJ} is
written as
\begin{equation}
\frac{D^2 J^{\mu}}{ds^2} + R^\mu_{\nu\rho\sigma} 
\frac{dq^\nu}{ds} J^\rho \frac{dq^\sigma}{ds} = 0~,
\end{equation}
and as already shown in Refs. \cite{Pettini,CasettiPettini}, 
in the case of Eisenhart metric it simplifies to
\begin{equation}
\frac{d^2 J^i}{dt^2} + \frac{\partial^2 V}{\partial q_i \partial q^j}
J^j=0~, \label{dintangEis}
\end{equation}
which is nothing but the usual 
tangent dynamics equation for standard Hamiltonians. 
The Lyapunov
exponents are usually computed evaluating the rate of exponential
growth of  $ J $ by means of a numerical integration of 
Eq. (\ref{dintangEis}) \cite{BenettinGS}.

In the particular case of {\em constant curvature} 
manifolds, Eq. (\ref{eqJ}) becomes very simple \cite{doCarmo}
\begin{equation}
\frac{D^2 J^\mu}{ds^2} + K \, J^\mu = 0~,
\label{eqJconst}
\end{equation}
and has bounded oscillating solutions $J \approx 
\cos(\sqrt{K}\, s)$ or
exponentially unstable solutions $J \approx \exp(\sqrt{-K}\, 
s)$ 
according to the sign of the constant sectional curvature 
$K$, which is
given by
\begin{equation}
K = \frac{K_R}{N-1} = \frac{{\cal R}}{N(N-1)}~,
\label{Kconst}
\end{equation}
where $K_R = R_{\mu\nu}\dot q^\mu \dot q^\nu$ is the Ricci 
curvature
and ${\cal R} = R^\mu_\mu$ is the scalar curvature; 
$R_{\mu\nu}$ is
the Ricci tensor. 
Manifolds with $K < 0$ are considered in abstract ergodic 
theory (see e.g. Ref. \cite{Sinai}).   
Krylov \cite{Krylov} originally
proposed that the presence of some negative curvature
could be the mechanism actually at work to make chaos in 
physical
systems, but in realistic cases the curvatures are neither 
found constant
nor everywhere negative, and the straightforward approach
based on Eq. (\ref{eqJconst}) does not apply. This is the 
main reason why
Krylov's ideas remained confined to abstract ergodic theory 
with few exceptions. 

In spite of these major problems, some approximations on Eq. 
(\ref{eqJ})
are possible even in the general case.
The key point is that negative 
curvatures are not strictly 
necessary to make chaos, and that a subtler 
mechanism related to the {\em bumpiness} of the ambient 
manifold is  actually at work. 

Let us choose a geodesic frame (i.e. a reference frame which is parallel 
transported along a geodesic; as a consequence, $D/ds \equiv d/ds$), 
and project Eq. (\ref{eqJ}) on a direction determined by the unit vector $u$:
\begin{equation}
\frac{d^2}{ds^2} \langle J,u \rangle + \langle R(\dot\gamma,J)\dot\gamma,u
\rangle = 0~.
\label{eqJproj}
\end{equation}
If the system is chaotic, $J$ grows exponentially with growth rates given
by the Lyapunov exponents, and if $u$ is the direction
corresponding to the largest Lyapunov exponent $\lambda$, after a finite
(proper) time $s$ the components of $J$ along the other directions will 
become negligible compared to that along $u$, thus we find $J \approx \psi u$.
Equation (\ref{eqJproj}) is thus rewritten approximately as a scalar 
Hill equation for $\psi$,
\begin{equation}
\frac{d^2\psi}{ds^2} + K(s)\,\psi = 0~,
\label{eqHill}
\end{equation}
where $K(s) = K(\dot\gamma,u)$ is the sectional curvature of the geodesic 
plane spanned by the directions $\dot\gamma$ and $u$ and is no longer
a constant, but a fluctuating function taking mostly {\em positive} values
(in some cases like the FPU model $K$ is strictly positive) whence the 
solutions of Eq. (\ref{eqHill}) can be subject to 
{\em parametric instability}. Curvature fluctuations can
produce chaos even if no negative curvature is experienced by the geodesics. 
As the sectional curvature is no longer constant, $K_R$ and
${\cal R}$ are respectively averages of $K$ over the 
direction
of $J$ and over both the direction of $J$ and
the direction of the reference geodesic in the latter. 
Equation 
(\ref{Kconst}) no longer holds, nevertheless it is a first 
order
approximation to which an estimate of the  curvature 
fluctuations
can be added to obtain a stochastic model of $K(s)$ 
independent of the
dynamics of the system \cite{Gaussmodel}. This model 
leads to an analytical estimate of the Lyapunov exponent
$\lambda$, which is correct (at least for the FPU model)
in the limit $N \rightarrow \infty$.

Up to this point the results are independent of the choice of the
metric. Specializing to the
Eisenhart arc-length parametrization Eq. (\ref{eqHill}) is rewritten in
terms of the time $t$,
\begin{equation}
\ddot\psi + K(t)\,\psi = 0~,
\label{eqHill(t)}
\end{equation}
where a dot stands for a time derivative and $K(t) = K(s\sqrt{2C_1^2})/2C_1^2$.
The stochastic model of $K(t)$ is given by
\begin{equation}
K(t) = \langle k_R \rangle + 
\langle \delta^2 k_R \rangle^{1/2}\eta(t)~,
\label{Kstoc}
\end{equation}
where $k_R= K_R/N$, $\langle \cdot \rangle$ stands for an average
taken along a geodesic, which, for systems in thermal equilibrium, 
can be substituted with a statistical average taken with respect to a
suitable probability measure (e.g.
the micro-canonical or the canonical measure); $\eta(t)$ is a  stationary 
$\delta$-correlated  Gaussian stochastic process with zero mean
and variance equal to one. Using Eisenhart metric, 
and for standard Hamiltonians, the non-vanishing components of the
Riemann tensor are $R_{0i0j} = \partial_{q_i} \partial_{q_j} V$,
hence the Ricci curvature has the remarkably simple form 
\begin{equation}
k_R = \frac{1}{N} \nabla^2 V~,
\label{k_R}
\end{equation}
where $\nabla^2$ is the Euclidean Laplacian operator. 
Equation (\ref{eqHill(t)}) becomes
a stochastic differential equation, i.e. the evolution equation
of a random oscillator \cite{VanKampen}. It is worth noticing
that Eq. (\ref{eqHill(t)}) is no longer dependent on the dynamics,
since the random process depends only on statistical averages.
The estimate of the Lyapunov exponent $\lambda$ is then obtained 
through the evolution of the second moments of the solution of 
(\ref{eqHill(t)}) as
\begin{equation}
\lambda = \lim_{t \to \infty} \frac{1}{2} \log 
\frac{\psi^2(t)+\dot\psi^2(t)}{\psi^2(0)+\dot\psi^2(0)}~.
\end{equation}
As shown in Ref. \cite{Gaussmodel}, this yields the following expression
for $\lambda$:
\begin{equation}
\lambda\,(k,\delta_k,\tau) = 
\frac{1}{2}\left(\Lambda-\frac{4 k}{3\Lambda}\right)~, 
\label{formula}
\end{equation}
where
\begin{mathletters}
\begin{equation}
\Lambda =
\left( \delta_k^2\tau + \sqrt{\frac{64 k^3}{27}+
\delta_k^4\tau^2}~\right)^{1/3}~, 
\end{equation}
\begin{equation}
\tau=\frac{\pi\sqrt{k}}{2 \sqrt{k(k + \delta_k)} + \pi\delta_k} ~;
\end{equation}
\end{mathletters}
in the above expressions $k$ is the average Ricci curvature 
$k = \langle k_R \rangle$ 
and $\delta_k$ stands for the mean-square fluctuation of the Ricci
curvature, $\delta_k = \langle \delta^2 k_R \rangle^{1/2}$.

The advantages in using the geometric approach to Hamiltonian chaos are
thus evident. In fact, it is possible to give reliable estimates of
the Lyapunov exponent without actually computing the time evolution
of the system: the estimate (\ref{formula}) of $\lambda$ depends only
on statistical averages which can be either computed analytically in
some cases (for instance in the case of the FPU model \cite{Gaussmodel})
or, in general, extracted from a Monte Carlo simulation, as it is the case
of the model to be studied in the present work. 

The behavior of the average geometric observables as the
control parameter (e.g. the energy density or the temperature) is varied
conveys an information which goes beyond the possibility of computing the
Lyapunov exponent. The dependence of the average Ricci curvature on the
energy density has already been used in Ref. \cite{CasettiPettini}
to give an operational definition of the SST which allows its 
computation in the thermodynamic limit, showing the stability of the
threshold in this limit for the FPU model. In fact, it is easy to
find that for a harmonic chain $\langle k_R \rangle$ is constant as 
the energy density is varied, and this is a common feature of other
integrable models (e. g. the Toda chain \cite{CasettiPettini}).
A computation of $\langle k_R \rangle(e)$ 
at constant volume (length) for the FPU
chain shows that the average Ricci curvature exhibits two well-defined
asymptotic behaviors, $\langle k_R \rangle(e) = \text{const.}$ as 
$e \to 0$ (harmonic limit), and $\langle k_R \rangle(e) = e^{1/2} $ as 
$e \to \infty$. The crossover between the two asymptotic curves occurs
at a value of $e$, $e_c$, which can be interpreted as a geometric estimate
of the SST. Such a geometric estimate is in very good agreement with
estimates based on other methods.

Moreover, one can look at the random oscillator equation
(\ref{eqHill(t)}) as an effective Jacobi equation for a geodesic flow
on a surface
$M$ whose Gaussian curvature is given by the random process $K(t)$. As long
as nonlinear coupled oscillators are considered, the average Ricci curvature
is positive, hence $M$ can be regarded as a sphere with a fluctuating radius.
In the limit of vanishing fluctuations, one recovers the bounded evolution
of the Jacobi field associated with integrable dynamics. Chaos suddenly
appears as curvature fluctuations are turned on, nevertheless it 
it will be ``weak'' as long as $\delta_k \ll k$, i.e. as long as
$M$ can be considered as a weakly perturbed sphere. On the contrary
as the size of curvature fluctuations becomes of the same order
of the average curvature, $\delta_k \approx k$, 
$M$ can no longer resemble a sphere, and the dynamics will no longer ``feel''
the integrable limit. Hence we expect the dynamics to be strongly chaotic.
This is by no means a deep explanation of the
existence of weakly and strongly chaotic regimes in Hamiltonian dynamics.
Nevertheless it shows how the simple geometric concepts which enter
the Riemannian description of Hamiltonian chaos, besides providing 
effective computational tools, are also useful in helping one's physical
intuition with images and analogies which would be difficult to find
elsewhere.

\section{The model and the geometric observables}
\label{model}

The system studied in the present work is 
a crystal of $n$ atoms of mass $m$ moving in three 
dimensions and interacting through a pairwise 
central potential $v(r)$. Its Hamiltonian is
\begin{equation}
{\cal H} =\frac{1}{2m} \sum_{i=1}^n {\bf p}_i^2   +  V({\bf X}) 
\label{hamil}
\end{equation}
where 
\begin{equation}
V({\bf X})=
\frac{1}{2} 
\sum_{i,j=1}^n 
v(|{\bf x}_i -{\bf x}_{j} |)~, 
\label{pot}
\end{equation}
The geometric observables which, within the approximations described in 
the previous Section, are relevant to
the dynamical instability of the system, are the average Ricci curvature
of $(M\times{\Bbb R}^2,g_E)$ and its fluctuations. 
In the case of a pairwise interaction potential
the Ricci curvature turns out to be (see Eq. \ref{k_R}):
\begin{equation}
k_R =
\frac{1}{N} 
\sum_{i=1}^n\sum_{j_i=1}^{12}
v''(|{\bf x}_i -{\bf x}_{j_i} |) + 2 \frac{v'(|{\bf x}_i -{\bf x}_{j_i} |)}
{|{\bf x}_i -{\bf x}_{j_i} |}    ~,
\label{kR3d}
\end{equation}
where it should be noticed that $N=3n$ as it represents the number of 
degrees of freedom.
The quantities which have to be determined are 
\begin{mathletters}
\label{averages}
\begin{equation}
k = \langle k_R \rangle 
\end{equation}
\begin{equation}
\delta_k^2 = \langle \delta^2 k_R \rangle = 
\frac{1}{N} \left(
\langle k_R^2 \rangle
- \langle k_R \rangle^2\right)~.
\end{equation}
\end{mathletters}
The probability measure which is usually employed for computational purposes  
is the canonical distribution so that the
statistical averages of Eqs.(\ref{averages}) can be written as
\begin{equation}
\langle  f \rangle (\beta)=Z_C^{-1} \int 
d{\bf X}\, 
f({\bf X})\,e^{-\beta V({\bf X})} ~, \label{int3dGibbs}
\end{equation}
where $Z_C$ is the configurational part of the canonical partition
function,
\begin{equation}
Z_C = \int 
d{\bf X}\, e^{-\beta  V({\bf X})} ~. \label{ZC3d}
\end{equation}
As a matter of fact in the canonical statistical ensemble the role of
control parameter is played by
the inverse temperature $\beta=1/k_B T$ where $k_B$ 
is the Boltzmann's constant.
However, in the thermodynamic limit ($N \to \infty$), the micro-canonical averages
can also be obtained from the canonical ones. 
As long as the mean value
$k = \langle k_R \rangle$ is concerned, the canonical and 
microcanonical averages differ only by a ${\cal O}(1/N)$ correction. 
Hence in the thermodynamic limit 
\begin{mathletters}
\begin{equation}
\langle k_R \rangle \left(e \right) = \langle k_R \rangle 
\left( \beta \left( e \right) \right)~,
\end{equation}
\begin{equation}
e \left(\beta \right) =  \frac{1}{2\beta} - 
\frac{1}{N}\frac{\partial}{\partial\beta}\left[ \log Z_C(\beta) \right]~. 
\label{e(beta)}
\end{equation}
\end{mathletters}
As regards $\delta_k^2 = \langle \delta^2 k_R \rangle$, one must keep in mind 
that fluctuations depend on the statistical ensemble. In fact the
difference between canonical and microcanonical fluctuations does not
vanish in the thermodynamic limit and the relation between these two
quantities is, according to Ref. \cite{LPV}, 
\begin{mathletters}
\begin{equation}
\langle \delta^2 k_R \rangle \left(e \right) = \langle \delta^2 k_R \rangle 
\left( \beta \left( e \right) \right) + F \left( \beta 
\left( e \right) \right) ~,
\end{equation}
\begin{equation}
F(\beta) = -\frac{k_{B}\beta^2}{c_v} 
\left(  \frac{\partial \langle k_R \rangle}{\partial \beta} \right)^2~,
\end{equation}
\end{mathletters}
where $e \left(\beta \right)$ is given again by Eq.(\ref{e(beta)}) and 
$c_v$ is the specific heat at constant volume.
Thus the micro-canonical fluctuations can be obtained from the canonical
ones provided that the values of the specific heat are known.

\section{Results and discussion}
\label{results}

The geometric observables described in the previous Sections 
have been evaluated 
for a Lennard-Jones face-centered cubic (fcc) crystal whose Hamiltonian 
is given by Eqs.(\ref{hamil}) and (\ref{pot})  
with a pair interaction potential which reads
\begin{equation}
v(r) = 4\varepsilon \left[ \left( \frac{\sigma}{r} \right)^{12} -
\left( \frac{\sigma}{r} \right)^6 \right]~. \label{LJ3d}
\end{equation}
Through an appropriate choice of the free parameters $m$ (mass), 
$\varepsilon$ and $\sigma$,
this simple model is able to take account of most thermodynamical
properties of  rare gas solids. 

The statistical averages of Eqs. (\ref{averages})
have been caculated by means of a standard
canonical Monte Carlo algorithm where a simulation box of $n=256$
particles subjected to periodic boundary conditions has been used.
In all simulations only nearest neighbors interactions have been 
dynamically taken into
account; the contributions of the interactions beyond the nearest neighbors
shell has been considered in a ``static approximation'' 
in which the istantaneous
relative positions of the atoms are replaced by their equilibrium
values. Apparently this procedure does not affect the evaluation of
the geometrical observables; its advantage resides in the fact that
allows one to employ the all neighbors parameters $\varepsilon$
and $\sigma$  which give a reasonnable representation
of the real pair potential and the equation of state of rare gas
solids \cite{RGS}.   

All data reported here are given in dimensionless form by reducing
them with respect to the ``natural units'' of the model; namely the
reduced energy is measured in unit of $\varepsilon$ and the Ricci curvature
in unit of $\varepsilon/\sigma^2$ as well as its mean-square fluctuation.

We performed two distinct series of simulations. In the first series
the density $\rho$ of the crystal has been kept constant
in order to compare the qualitative behaviour of the geometric observables
with the known results for one-dimensional systems.  
In fact, computations of chains
of anharmonic oscillators have been
performed at constant length \cite{Pettini,CasettiPettini,1d}.
The numerical outcomes of this series of simulations 
are reported in Figs. 1-3 where $k$, $\delta_{k} / k$ and the estimate
$\lambda$ of the Lyapunov exponent computed according to Eq. (\ref{formula}) 
are plotted versus the energy per degree of freedom of the system. 
The additional data
appearing in these figures refer to computations where the interaction
potential (26) has been expanded up to the fourth order in the Taylor series.
This procedure allows us to perform simulations at higher energies in order to 
provide a better representation of the crossover energy  region.
The results reported in Fig. 1 show that the SST, or crossover
energy,  can be located around $e_c \simeq 0.15$. This estimate of $e_c$ 
is confirmed by the results reported in Fig. 2, for 
Fig. 2 shows that in correspondence
of this threshold the ratio $\delta_k/k$ deviates remarkably from the
low-energy behaviour $\delta_k/k \propto e$ and tends to saturate towards
$\delta_k/k \simeq 1$ which implies that the manifold becomes highly
anisotropic, and thus we expect the dynamics to enter the strongly chaotic
regime. As a consequence of the behaviour of $k$ and $\delta_k$, 
the geometric estimate
$\lambda$ of the Lyapunov exponent shows a sharp crossover between
two different power laws. It is worth recalling that the quantity here
reported is not the ``true'' Lyapunov exponent but an estimate which has
anyhow proven to be extremely accurate in other systems \cite{Gaussmodel}.
 
The aim of the second series of simulations is mainly to determine whether
the averages of the geometric observables are affected by a change in the
density of the sample.
In principle one expects such a change because
the curvature properties depend on space derivatives, and a change of the
density induces a change in the length scale. 
Moreover, it is interesting to test the expectation that the crossover in the 
relaxation behaviour of a Xe crystal  recently observed
in numerical simulations 
\cite{3deuro}, is related with some major change in the global properties 
of the dynamics (i.e. the SST) which in turn can be detected by the 
geometric observables under investigation in this paper. 
Hence it is interesting to compute the statistical averages of 
Eqs. (\ref{averages})
using settings which are close to those 
used in the molecular dynamics simulations reported in \cite{3deuro}.
In particular the results shown in Figs. 3-6 refer to
the values of the density 
given by the empirical equation of state of solid Xe 
($\varepsilon/k_{B}=228.6^\circ$ K, $\sigma$ = 3.959 \AA)
\cite{RGS}.

The results reported in Figs. 3-6 show that the qualitative behaviour
of the curvature fluctuations (Fig. 5) and of the 
theoretical
estimate of the Lyapunov exponent $\lambda$ (Fig. 6) is the same as
in the constant density case, while the behaviour of the average
curvature (Fig. 4) is dramatically altered: instead of a crossover 
between two asymptotic regimes we have here a neat maximum of $k$
at $e = e_c$.
It is remarkable that the value
of $e_c$ -- as estimated through the behaviour of $\delta_k$ or $\lambda$
or by the position of the maximum of $k$ -- remains the same as
in the constant density case, and
corresponds to a temperature $T\simeq 0.15 \varepsilon/k_B$ , i.e. 
occurs  at a temperature of  physical relevance for 
the thermodynamics of this system.
Moreover, the value of $e_c$ is in excellent agreement with the value
of the crossover energy  estimated via the nonequilibrium dynamics 
\cite{3deuro}. 

Let us now briefly comment about the problem of the relevance
of quantum effects for the results presented in this Section. As already
stated in the Introduction, our analysis is a completely classical one
so that the significance of our results  depends on
the reliability of the classical approximation of the dynamics in the 
range of temperatures investigated.
The strength of the quantum behavior of the interaction
is ruled by a coupling parameter $g=\hbar\omega_{LJ}/\varepsilon$ that is  
the ratio
between the typical vibrational quantum energy $\hbar\omega_{LJ}$ 
calculated in the harmonic 
approximation and the binding energy $\varepsilon$. Values $g\ll 1$ 
denote that quantum effects may be neglected
in a wide range of temperatures. However, as observed in Ref. \cite{pimcrgs},
the failure of the classical approximation becomes
quite evident for all rare gas solids at 
measurable  temperatures.
The detailed evaluation of the influence of quantum effects for the
determination of the threshold energy $e_c$ is out of the purposes of 
this paper. We present here results for solid Xenon where, due to the
small coupling parameter ($g=0.106$), the relevance of quantum effects
in our discussion, if any, is certainly not decisive.

\section{Conclusions}
\label{conclusions}

We have presented here a detailed description of how a differential
geometric approach to Hamiltonian chaos can be used for determining
the  existence of the SST in a realistic system. We have discussed
the connection between the SST and the geometrical observables which
have been investigated in this paper. We have shown how  standard
methods of classical statistical mechanics can be used for our 
computational purposes. We have finally applied the framework here
developed to a Lennard Jones crystal modeling solid Xenon. The 
crossover energy region of this system has been clearly detected and the value
of the energy threshold has turned out to be in excellent agreement with
the numerical results presented in a recent paper \cite{3deuro}. 
The possibility of setting up an experiment of this system for determining 
the physical consequences of the transition from weak and strong
chaos regime is under investigation.

\acknowledgments

It is a pleasure to thank Roberto Livi and Marco Pettini for 
enlightening discussions and for their interest in our work.

\begin{figure}
\caption{Average Ricci curvature $k$ vs the energy
         per degree of freedom at constant density ($\rho=1.0$ in natural
         units). Open circles: data of computations where the 
         interaction potential (26) has been expanded up to the fourth order
         in the Taylor series;
         solid circles: data obtained with the ``full'' interaction potential
         (26). The two dotted lines represent the estimated low- and
         high-energy behaviours of $k$ 
         helping the identification of $e_c$; the line interpolating
         the high-energy data has been obtained by fitting a law
         $k \approx  \alpha + \beta \, e^{1/2}$, with fitting
         parameters $\alpha$ and $\beta$, to the high-energy data. }
\label{fig1}
\end{figure}

\begin{figure}
\caption{Normalized mean-square fluctuation of the Ricci curvature 
         $  \delta_{k} / k$ vs 
	 the energy per degree
	 of freedom at constant density. Symbols as in Fig. 1.
	 The dotted line is the linear behaviour.}
\label{fig2}
\end{figure}

\begin{figure}
\caption{Theoretical estimate of the Lyapunov exponent $\lambda$ 
	according to Eq. (\protect\ref{formula})
	versus the energy per degree of freedom
	 at constant density. Symbols as in Fig. 1. The dotted line is
	 the power law $e^2$.}
\label{fig3}
\end{figure}

\begin{figure}
\caption{Average Ricci curvature $k $ vs the energy
         per degree of freedom. Here, in each simulation we have used values 
         of the density according to the empirical equation of state of solid 
         Xenon.}         
\label{fig4}
\end{figure}

\begin{figure}
\caption{The same as Fig. 4 for the normalized means-square fluctuation of 
         the Ricci curvature $  \delta_{k} / k$. The dotted line is the
         linear behaviour.}
\label{fig5}
\end{figure}

\begin{figure}
\caption{The same as Fig. 4 for the estimate of the 
	Lyapunov exponent $\lambda$. The dotted line is the law $e^2$.}
\label{fig6}
\end{figure}

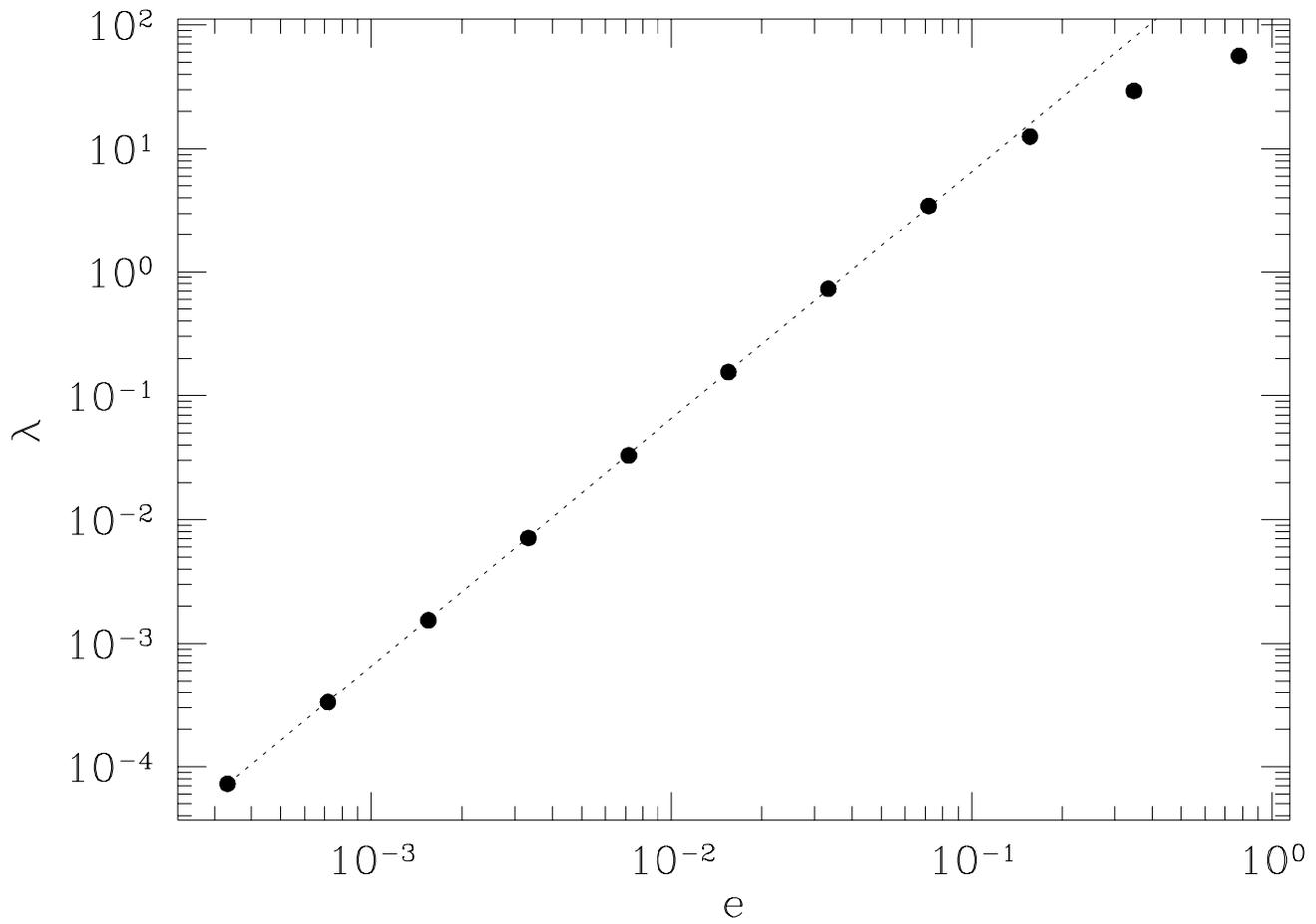
\end{document}